\documentclass[article,twocolumn,nofootinbib,prd]{revtex4}
\usepackage[letterpaper,left=.75in,right=.75in,top=.75in,bottom=1.in]{geometry}
\usepackage{graphicx,amsfonts,color,comment,amsmath,hyperref,float}
\usepackage{times}
\usepackage{amssymb}	
\usepackage{scrextend}

\usepackage{mathrsfs,amssymb}  
\usepackage{cancel}
\usepackage[normalem]{ulem}

\newcommand{\be}{\begin{equation}}
\newcommand{\ee}{\end{equation}}
\newcommand{\bea}{\begin{eqnarray}}
\newcommand{\eea}{\end{eqnarray}}

\begin{document}
\title{Direct Detection Experiments 
at the Neutrino Dipole Portal Frontier} 

\author{Ian M. Shoemaker}
\author{Jason Wyenberg}

\affiliation{Department of Physics, University of South Dakota, Vermillion, SD 57069, USA}

\date{\today}
\begin{abstract}

Heavy sterile neutrinos are typically invoked to accommodate the observed neutrino masses, by positing a new Yukawa term connecting these new states to the neutrinos in the electroweak doublet. However, given our ignorance of the neutrino sector, we should explore additional interactions such sterile neutrinos may have with the SM. In this paper, we study the dimension-5 operator which couples the heavy state to a light neutrino and the photon.  We find that the recent XENON1T direct detection data can improve the limits on this ``Neutrino Dipole Portal'' by up to an order of magnitude over previous bounds. Future direct detection experiments may be able to extend these bounds down to the level probed by SN1987A. 

\end{abstract}
\preprint{}


\maketitle
\section{Introduction}

The fact that neutrinos are massive is one of the key observational facts indicating that the Standard Model (SM) of particle physics is incomplete. Most models of neutrino masses posit new right-handed states which are singlets under the SM gauge groups. These neutral fermion singlets have been predominantly studied in connection with neutrino masses via the Neutrino Portal interaction, $\mathscr{L}  \supset NHL$, where $N$ is the singlet fermion, $L$ is the SM lepton doublet, and $H$ is the Higgs doublet.  For this reason, singlet fermions can play the role of a ``sterile'' neutrino (i.e. uncharged under the electroweak symmetry), and they mix with the left-handed neutrinos after the Higgs acquires a vacuum expectation value.  

However, the standard neutrino portal interaction may not be the predominant interaction these states have with the SM.  They may also interact via a ``Neutrino Dipole Portal'' interaction, which after electroweak symmetry breaking can be written as
\be 
\mathscr{L}_{{\rm NDP}}  \supset d \left(\bar{\nu}_{L}  \sigma_{\mu \nu} F^{\mu \nu} N \right) + h.c.,
\ee
where $F_{\mu \nu}$ is the electromagnetic field strength, $\sigma_{\rho\sigma}=\frac{i}{2}[\gamma_\rho,\gamma_\sigma]$, $\nu_{L}$ is the SM neutrino, and the coefficient $d$ with units of $(\rm{mass})^{-1}$ controls the strength of the interaction. Despite its simplicity and the wide interest in a ``sterile'' neutrino, this interaction has not received much attention. It has, however, been considered in the context of the MiniBooNE events~\cite{Gninenko:2009ks,Gninenko:2010pr,McKeen:2010rx,Masip:2011qb,Gninenko:2012rw,Masip:2012ke,Bertuzzo:2018itn}, and has also been been studied in the context of IceCube data~\cite{Coloma:2017ppo} and at the upcoming SHiP experiment~\cite{Magill:2018jla}. 

In this paper we will study the neutrino dipole portal (NDP) at the XENON1T direct detection experiment using their $\simeq 1$ ton-year exposure~\cite{Aprile:2018dbl}.  Despite not finding evidence of DM scattering, XENON1T is nearly at the level where they can start seeing events from solar neutrinos.  Many prior works have used neutrinos at direction detection experiments to study various beyond SM neutrino {interactions}~\cite{Pospelov:2011ha,Harnik:2012ni,Pospelov:2012gm,Pospelov:2013rha,Coloma:2014hka,Cerdeno:2016sfi,Dent:2016wcr,Bertuzzo:2017tuf,Dutta:2017nht,AristizabalSierra:2017joc,Gonzalez-Garcia:2018dep,Boehm:2018sux}.  At the few keV recoil energies of XENON1T, the Boron-8 (B8) neutrinos make the largest contribution, comprising $\sim 0.02$ background events in the 1 ton-year sample~\cite{Aprile:2018dbl}.  However, if neutrinos have additional interactions beyond EW forces, this rate could be larger and already detectable. To get an approximate idea of the sensitivity to the NDP we can compare the SM cross section, $d \sigma /dE_{R} \simeq G_{F}^{2} Q_{w}^{2} m_{N}/4 \pi$, with the NDP cross section,  $d \sigma /dE_{R} \simeq d^{2} \alpha Z^{2} /E_{R}$. 

\begin{figure}[b!]
  \centering
  \includegraphics[width=.9\linewidth]{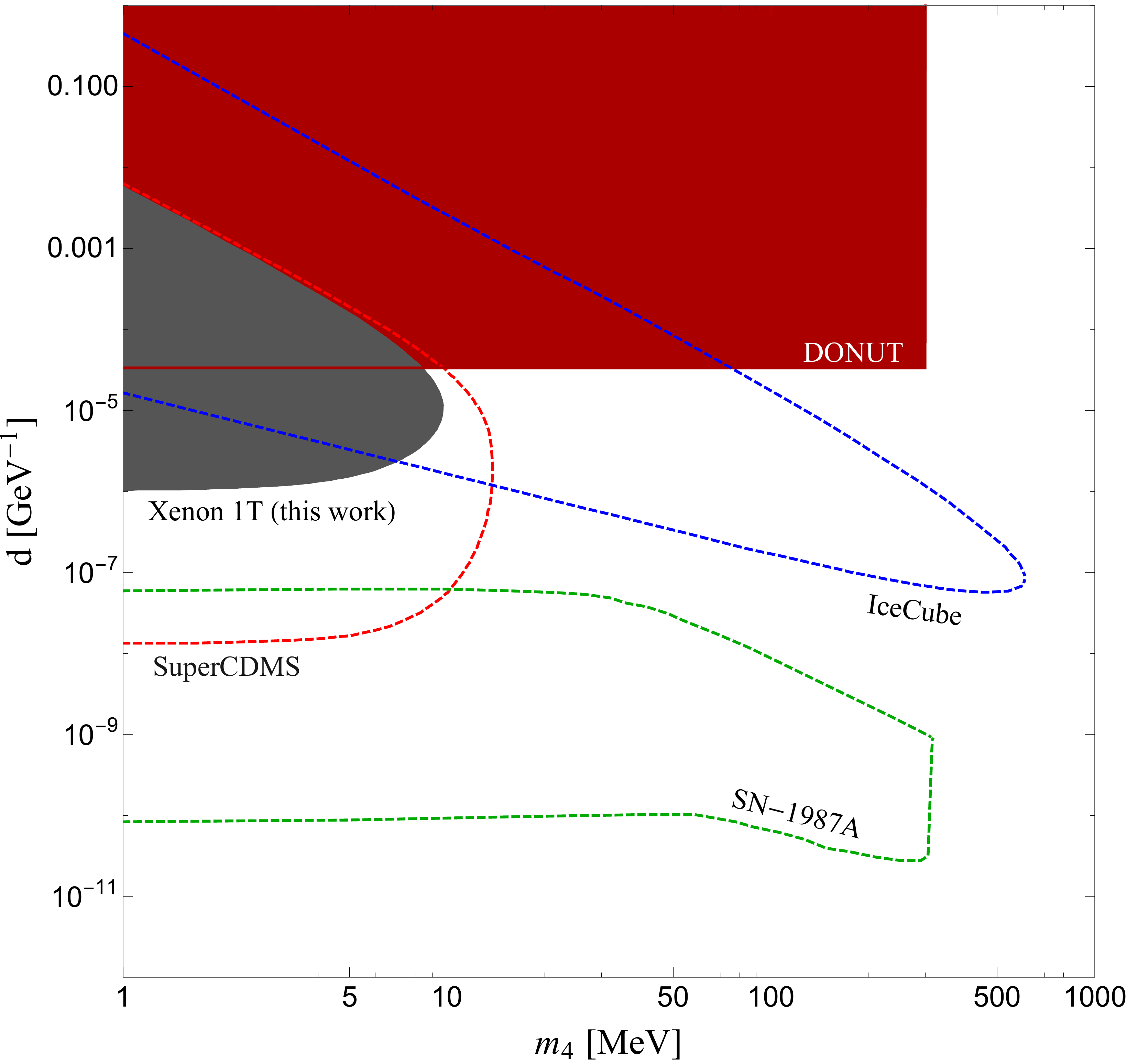}
  \caption{These are the expected sensitivity curves for the tau-flavored NDP based on the XENON1T data~\cite{Aprile:2018dbl} and a future SuperCDMS exposure.  The relevant previously published exclusion limits of SN1987A~\cite{Magill:2018jla}, IceCube~\cite{Coloma:2017ppo}  and DONUT~\cite{Schwienhorst:2001sj,Coloma:2017ppo} are also shown.}.
  \label{ExclusionCurves}
\end{figure}

Thus to achieve $\sim 1$ event at XENON1T we would very roughly expect 
\be d\simeq \sqrt{50 \frac{G_{F}^{2}Q_{w}^{2}m_{N}E_{R}}{4\pi \alpha Z^{2}}} \sim 10^{-6}~{\rm GeV}^{-1}~\sqrt{\frac{E_{R}}{{\rm keV}}},
\label{eq:est}
\ee
where the factor of 50 comes from needing a 50-fold increase in the SM cross section for the ``neutrino floor'' to presently be detectable.  We expect the estimate in Eq.~\ref{eq:est} to be valid up to singlet fermion masses of order Boron-8 energies, $m_{4} \sim E_{\nu} \sim 10$ MeV.  Although the above estimates are simplistic, they provide us with ample motivation to carry out a more complete analysis. Indeed, a dipole strength at the $d\simeq 10^{-6}~{\rm GeV}^{-1}$ level is competitive with a variety of known constraints on the NDP~\cite{Masip:2012ke,Coloma:2017ppo,Magill:2018jla}. We summarize our main findings in Fig.~\ref{ExclusionCurves} which demonstrate that XENON1T already provides the leading constraints up to 10 MeV masses, and future high-exposure/low-threshold direct detection can improve the bounds down to the SN1987A region.  

The remainder of this paper is organized as follows. In Sec.~\ref{sec:rate} we compute a realistic spectrum of nuclear recoil events at XENON1T from solar neutrinos mediated by the NDP interaction. We find there that the new state will typically decay outside the detector after it is produced and that an incoming neutrino is unlikely to undergo much up-scattering in the Earth prior to arrival at the detector. In Sec.~\ref{sec:future} we look at what improvements can come in the near term, focusing on a future run of SuperCDMS. Then in Sec.~\ref{sec:discuss} we discuss the nature of models giving rise to the NDP, along with potential ways in which future work could refine and extend the analysis carried out here.

\section{Neutrino Transition Magnetic Moment Rates}
\label{sec:rate}
\subsection{Mechanism}
%

\begin{figure}[t!]
  \centering
  \includegraphics[width=.7\linewidth]{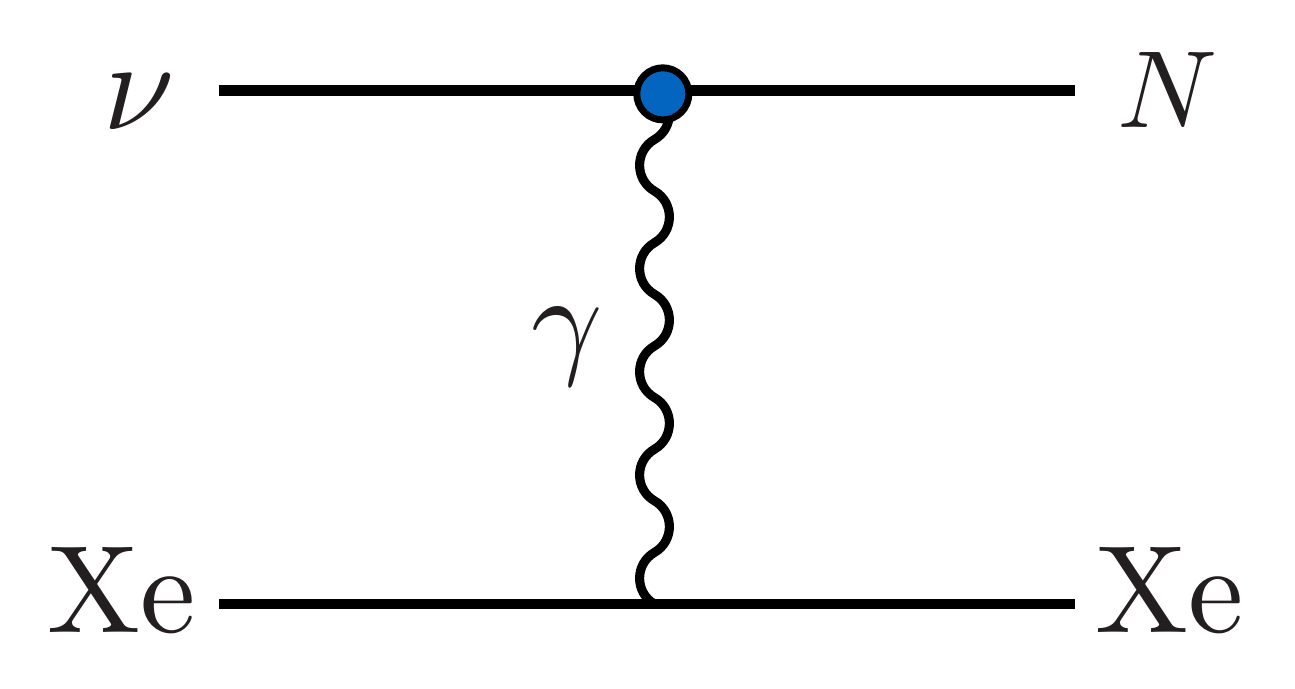}
  \caption{The neutrino dipole portal $\mathscr{L}_{{\rm NDP}}  \supset d \left(\bar{\nu}_{L}  \sigma_{\mu \nu} F^{\mu \nu} N\right)$ allows for a neutrino to up-scatter off a nucleus to a heavy neutral lepton state $N$. This produces a distinctive recoil spectrum, while the newly produced heavy neutral lepton decays outside the detector.}
  \label{fig:diagram}
\end{figure}

Incoming solar neutrinos may up-scatter to a heavy state $N$ through the NDP operator via the process shown in Fig.~\ref{fig:diagram}. The coherent cross section for neutrino-nucleus scattering via a NDP reads
\begin{widetext}
\bea
\label{eq:magneticmomentcrosssection}
\frac{d\sigma_{\nu n\rightarrow Nn}}{dE_R} &=& d^2\alpha Z^2 F^2(E_R)\Bigg[\frac{1}{E_R}-\frac{m_{4}^2}{2E_{\nu}E_{R}m_N}\Bigg(1-\frac{E_{R}}{2E_\nu}+\frac{m_N}{2E_\nu}\Bigg)-\frac{1}{E_\nu}+\frac{m_{4}^4(E_{R}-m_N)}{8E_\nu^2E_{R}^2m_N^2}\Bigg]
\eea
\end{widetext}
where $Z$ is the atomic number, $E_{R}$ is the nuclear recoil energy, $E_\nu$ is the incident neutrino energy, $m_N$ is the mass of the target nucleus, $m_{4}$ is the mass of the heavy sterile neutrino, and $F(E_{R})$ is the nuclear form factor.

\subsection{Event Rates in Xenon1T Detector}

Neutrino scattering at dark matter direct detection experiments has been widely studied~\cite{Cabrera:1984rr,Monroe:2007xp,Strigari:2009bq,Gutlein:2010tq,Billard:2013qya,Ruppin:2014bra,Dent:2016iht,OHare:2016pjy,Cerdeno:2016sfi,Essig:2018tss,Wyenberg:2018eyv} and depending on the range of energies may include contributions from solar, atmospheric and the diffuse supernova background. In fact, the original DM direct detection proposal from Goodman and Witten~\cite{Goodman:1984dc} was an extension of the detection method for solar and reactor neutrinos via neutral currents by Drukier and Stodolsky~\cite{Drukier:1983gj}. 

\begin{figure}[b!]
  \centering
  \includegraphics[width=1\linewidth]{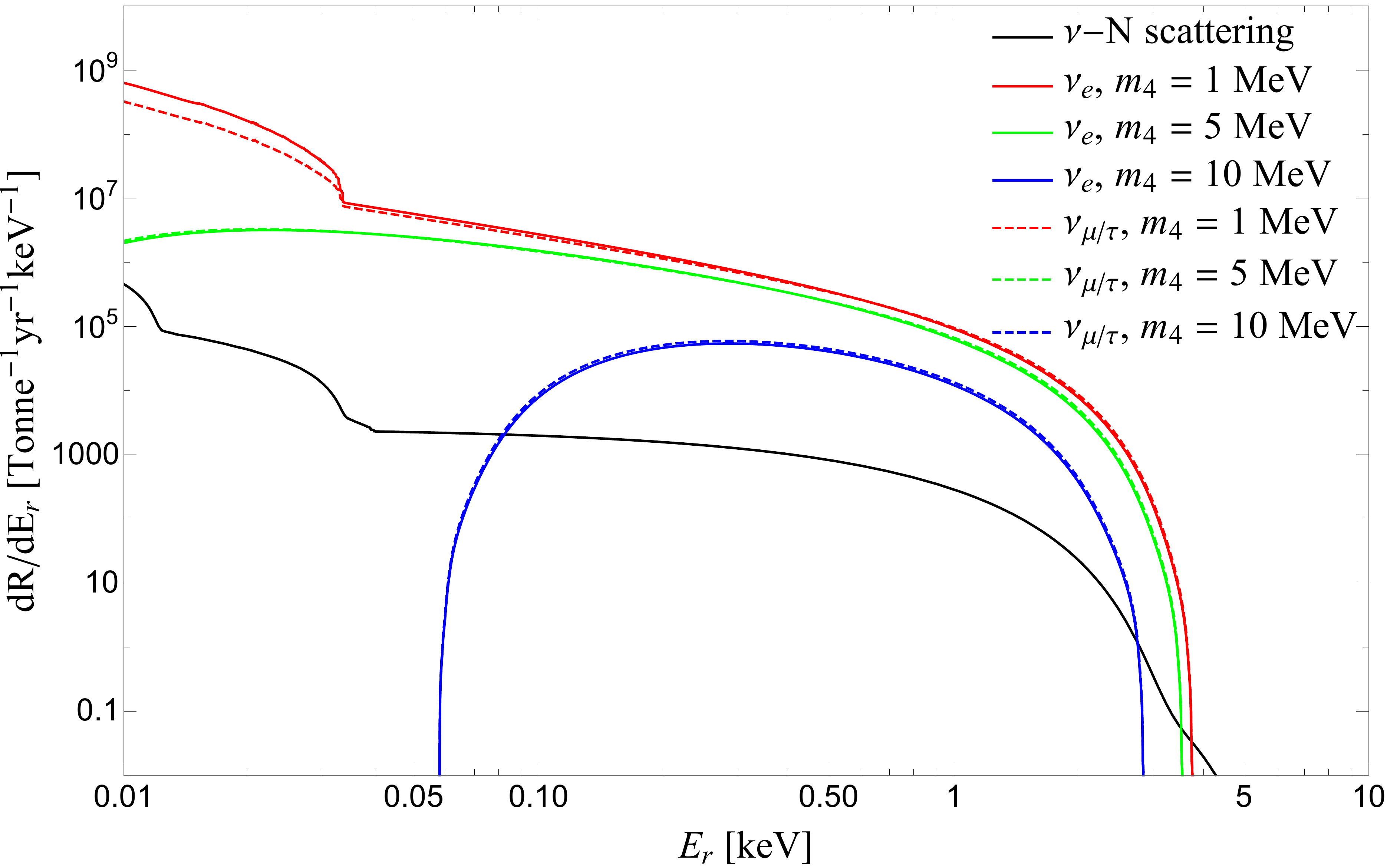}
  \caption{Event rates of nuclear scattering via a NDP in the Xenon 1T detector with $d=10^{-6}\text{ GeV}^{-1}$ for masses $m_{4} = 1$ MeV (red), $m_{4} = 5$ MeV (green), and $m_{4} = 10$ MeV (blue).  Also shown is the standard model $\nu$-nucleus scattering rate (black curve). }
  \label{Rates}
\end{figure}

The nuclear recoil spectrum of neutrino induced scattering events can be computed via:
\be
\label{eq:magneticmomentrate}
\frac{dR^\alpha}{dE_R} = MT\times\int_{E_{\nu}^{min}}\frac{d\Phi_{\nu}^\alpha}{dE_{\nu}}\frac{d\sigma_{\nu n\rightarrow Nn}^\alpha }{dE_R}(E_{\nu},E_{R}\big)dE_{\nu},
\ee
where $\alpha$=$e,\mu,\tau$ indexes the neutrino flavor, $MT$ is the exposure of the XENON1T experiment (equal to 1.3 Tonne$\times$278.8 days), and $\Phi_\nu$ is the solar neutrino flux. The  function $E_\nu^{{\rm min}}(E_{R}) $ is the minimum energy of the incident neutrino to up-scatter to the state of mass $m_{4}$ while producing a nuclear recoil $E_{R}$:
\be
\label{eq:minimumenergy}
E_\nu^{{\rm min}}(E_{R})=\frac{m_{4}^2+2m_NE_R}{2\Big[\sqrt[]{E_R(E_R+2m_N)}-E_R \Big]}.
\ee
The solar neutrino $^{8}B$ flux provides the dominant contribution to NDP nuclear scattering in the XENON1T experiment. We normalize the $^{8}B$ spectrum to the flux $\phi_{^{8}B} = 5.1 \times 10^{6}~{\rm cm}^{2}~{\rm s}^{-1}$~\cite{Abe:2010hy}. For illustration, in Fig.~\ref{Rates} we plot the expected recoil spectrum for a range of possible $\nu_{4}$ masses.  As one would expect, at low masses there is little dependence on the recoil spectrum, while at masses around $\sim 10$ MeV the rate starts to get very suppressed.

A detected nuclear recoil event will create a signal of $n$ photoelectrons (PEs) given by a poisson distribution with expectation value $\overline{n}$, given by:
\bea
\label{eq:expectedevents}
\overline{n}=E_{R}L_y\big(E_{R}\big)g_1.
\eea
$L_y\big(E_{R}\big)$ is the light yield as a function of $E_{R}$ as shown in~\cite{Akerib:2016mzi}, and detector photon gain is $g_1= 0.144\pm 0.007$~\cite{Aprile:2017iyp}. The event rate is then given by:
\bea
\label{eq:eventrate}
\frac{dR^\alpha}{dn}=\int\text{Eff}\big[E_{R}\big]~\frac{dR^\alpha}{dE_{R}}\times \text{Poiss}(n|\overline{n})~{dE_{R}},
\eea
where $\text{Eff}\big[E_{R}\big]$, the efficiency as a function of nuclear recoil energy, is given in Fig. 1 of Ref.~\cite{Aprile:2018dbl}. Finally, we model the total signal rate as:
\bea
\label{eq:signalrate}
\frac{dR^\alpha}{dS1}=\sum\limits_{n=1}^{\infty}\text{Gauss}\big(S1|n,\sqrt{n}\,0.5\big)\times\frac{dR^\alpha}{dn},
\eea
with the 0.5 factor coming from the uncertainty of the 1 PE bin size.
\subsection{Exclusion Curves}
The result of the XENON1T experiment excludes a portion of the $m_{4}, d$ NTMM parameter space. Following a Bayesian approach, with a signal $s$ and background $b$ an upper limit on $s$ can be determined as:
\bea
\label{eq:sup}
s_{up}=\frac{1}{2}F_{\chi^2}^{-1}[p,2(n+1)]-b,
\eea
where $F_{\chi^2}^{-1}$ is the inverse cumulative $\chi^2$ distribution, and $n$ is the number of observed events such that $2(n+1)$ is the number of degrees of freedom. The $p$ factor is given by the expression:
\bea
\label{eq:pfactor}
p=1-\alpha\big(F_{\chi^2}[2b,2(n+1)]\big),
\eea
where $\alpha$ is $1-CL$, and $CL$ is the confident limit \cite{Tanabashi:2018oca}. An alternative statistical analysis may employ the Likelihood Profile method~\cite{Aprile:2011hx} incorporating the binned energy data. A check of the calculated exclusion curve for $d$ for several values of $m_4$ showed nearly identical results between a rudimentary likelihood profile method and the $\chi^2$ approach employed here. 

\begin{figure}[t!]
  \centering
  \includegraphics[width=.9\linewidth]{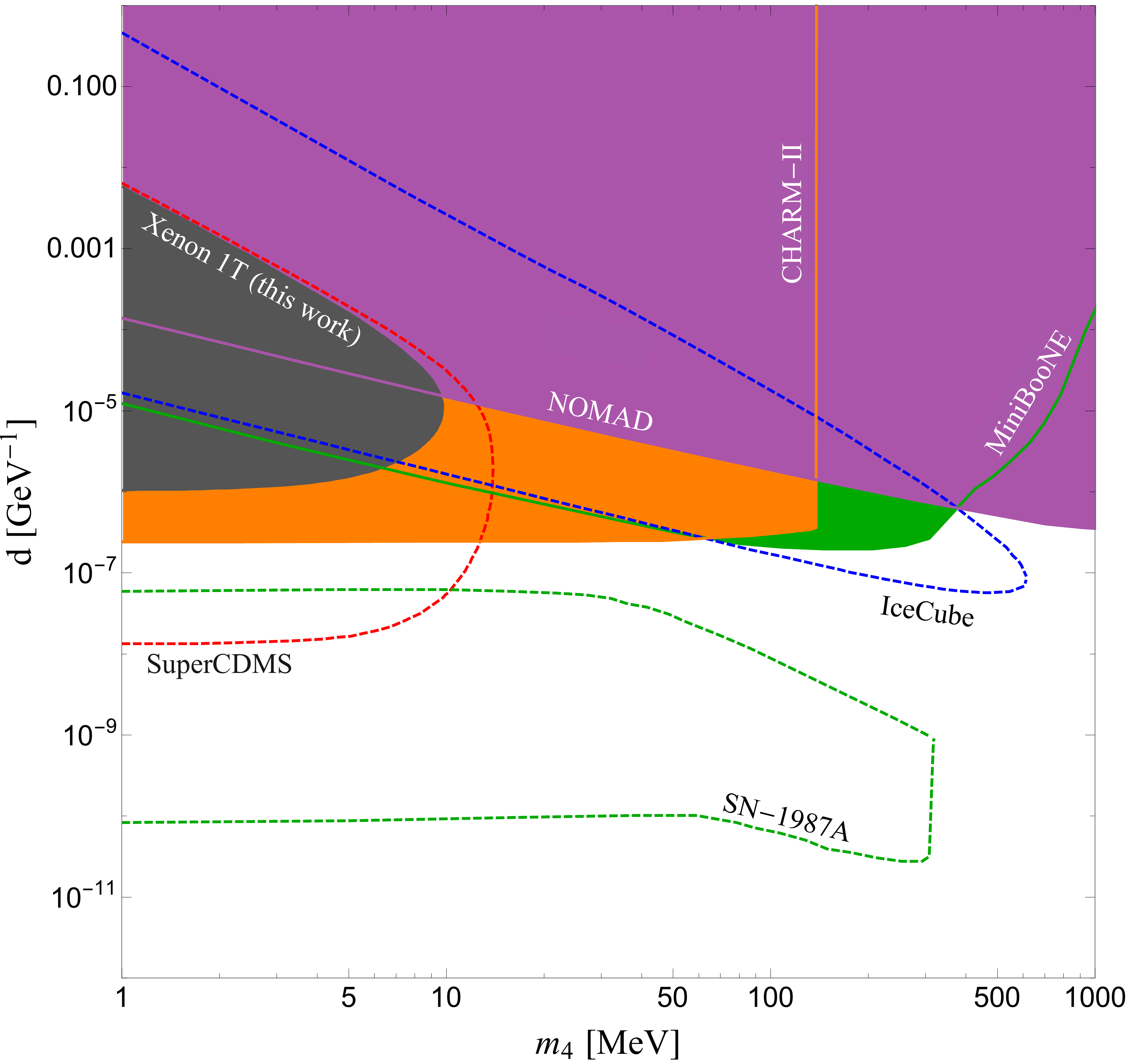}
  \caption{Expected sensitivity to muon-flavored NDP at XENON1T and a future SuperCDMS exposure. Included are bounds from NOMAD~\cite{Altegoer:1997gv,Coloma:2017ppo}, CHARM~\cite{Geiregat:1989sz,Coloma:2017ppo}, MiniBooNE~\cite{AguilarArevalo:2007it,Magill:2018jla}, IceCube~\cite{Coloma:2017ppo}, and SN1987A~\cite{Magill:2018jla}. }
  \label{fig:muon}
\end{figure}

For the XENON1T data, with 2 observed events and an expected background of 1.34 events, $s_{up}=6.53$. Figure \ref{ExclusionCurves} shows the 90\% confidence exclusion in the $(m_{4}, d)$ plane. Also shown are excluded regions from previously published results (see caption for details). For reference we also show the current and future direct detection sensitivity to the muon-flavored NDP in Fig.~\ref{fig:muon}.

\subsection{Up-scattering and Decay Considerations}

Notice that in principle the dipole interaction admits the possibility of $\nu \rightarrow N$ upscattering prior to the neutrino flux arriving at the detector. At minimum a neutrino traverses $\sim 1$ km to reach the underground detector. We will find that the process of up-scattering in the Earth is irrelevant for the parameters of interest. The total cross section for up-scattering is roughly estimated as~\cite{Magill:2018jla}
\be
\sigma_{\nu \rightarrow N} \simeq  \alpha Z^{2} |d|^{2} \times \log\left(\frac{4 E_{\nu}^{2}}{m_{4}^{4} R_{{\rm nuc}}^{2}}\right)
\ee
For an incoming solar neutrino with Boron-8 energies $E_{\nu} \simeq 10$ MeV while traversing a distance of 1 km through the Earth, we find that $d$ would need to be, 
\bea d &\simeq& \sqrt{\frac{1}{(1{\rm km})\, n_{\oplus}  \alpha Z^{2} \log\left(\frac{4 E_{\nu}^{2}}{m_{4}^{4} R_{{\rm nuc}}^{2}}\right)}} \\ 
& \simeq &0.14~{\rm GeV}^{-1}\nonumber
\eea
where we assumed that the dominant contribution to the terrestrial density is silicon. Dipole strengths this strong are already excluded by a number of independent probes including DONUT and IceCube ($\nu_{\tau}$ transitions)~\cite{Coloma:2017ppo}, and by CHARM-II, MiniBooNE and LSND~\cite{Magill:2018jla} ($\nu_{\mu}$ transitions).

{Shortly after being produced through up-scattering in the detector the $\nu_{4}$ state will eventually decay.  If this decay, $\nu_{4} \rightarrow \nu + \gamma$, happens inside the detector volume, the resultant photon could potentially cause the signal to be thrown away as background. Of course if the initial nuclear energy deposition in the up-scattering $\nu +$Xe $ \rightarrow \nu_{4} + $Xe is sufficiently far from the final decay of the $\nu_{4}$ one may still be able to perform a search using either only the nuclear recoil events or a dedicated search aimed at the spatially/temporally correlated nuclear and photon signals unique to the NDP model. Such an analysis is beyond the scope of this work, but could result in stronger bounds than the analysis performed here. }

To ensure that the nuclear recoil events we model have not been vetoed by the collaboration, we only exclude regions of parameter space where the newly produced $\nu_{4}$ state decays outside of the detector. To do this we impose a penalty factor to account for the possibility that the heavy state decays in the detector shortly after being produced. This penalty factor takes the form
\be 
P_{{\rm D}} = e^{-l_{{\rm det}}/\ell_{D}},
\label{eq:penalty}
\ee
where $l_{{\rm det}}$ is a characteristic length scale of the detector which we take to be 100 cm, and the boosted decay length for a $\nu_{4}$ with energy $E_{4}$ is
\be
 \ell_{D} \equiv \gamma \beta \tau \simeq \frac{16 \pi E_4}{d^{2} m_{4}^{4}} \sqrt{\left(\frac{E_{4}}{m_{4}}\right)^{2} -1}.
\ee
For the decay $N \rightarrow \nu + \gamma$ to occur inside within 100 cm we would need $d \gtrsim 4 \times 10^{-4}{\rm GeV}^{-1}$ for $E_4 \simeq 10$ MeV and $m_{4} = 5$ MeV.

\section{Future Direct Bounds: SuperCDMS ultra-low thresholds}
\label{sec:future}

The SuperCDMS collaboration~\cite{Agnese:2018col} anticipates detection of low energy nuclear recoil events which could exclude values of $d$ several orders of magnitude lower than the Xenon 1T data. For an order of magnitude estimate, a 1-ton year exposure of the SuperCDMS experiment was modeled. Using the published efficiency curve of the SuperCDMS experiment~ \cite{Agnese:2017jvy} (see their Fig. 4), the estimated background from solar neutrino nuclear recoils is $\sim 460$ events. For a $3\sigma$ discovery signal above the statistical uncertainty, $\sim 65$ signal events are required. Figure \ref{ExclusionCurves} includes a projected exclusion curve of the SuperCDMS experiment.

\begin{figure}[t!]
  \centering
  \includegraphics[width=1\linewidth]{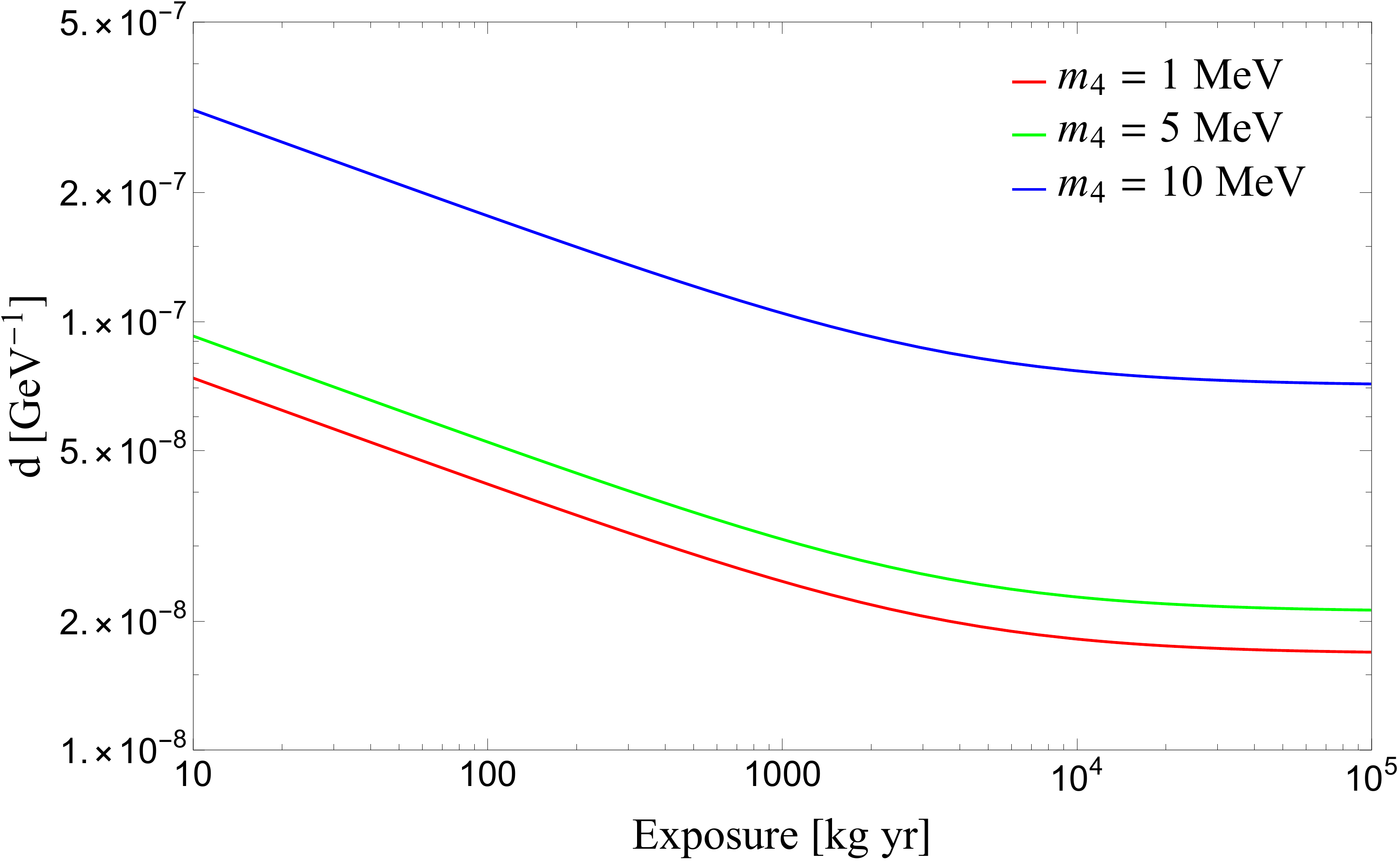}
  \caption{Anticipated sensitivity of SuperCDMS as a function of exposure, eventually plateauing from systematic uncertainties. }
  \label{SuperCDMSExposure}
\end{figure}

Worth noting is the fact that the SuperCDMS collaboration may soon be probing the neutrino floor. For exposures much higher than 1-ton year, the detector reach plateaus and cannot probe any lower values of $d$. This is shown in Figure \ref{SuperCDMSExposure} as the neutrino background systematic uncertainty obscures a potential signal from the neutrino magnetic moment.

Eventually $pp$ solar neutrinos may contribute to the neutrino background at direct detection experiments. This would require very low-energy thresholds and would result in a large enhancement of the event rate. For the mechanism of the neutrino transition magnetic moment discussed in this paper, we estimate that a future detector would require a recoil energy threshold below $\sim 1$ eV to observe the $pp$-flux, and only if $m_4$ is below $\sim 100$ keV.

The upper limit of $m_4$ that could be excluded by the Xenon 1T experiment is $\sim 10$ MeV due to the maximum incoming energy of the relevant solar neutrino flux. For the exposure of the Xenon 1T data, this maximum energy is that of the solar neutrino B8 flux. To probe higher values of $m_4$, the atmospheric neutrino flux could be employed, but this requires an exposure of $\sim 10^7$ ton-year for a Xenon based experiment.

\section{Model Discussion}
\label{sec:discuss}

{Throughout the paper we have assumed that the NDP term, $\mathscr{L}_{{\rm NDP}}  \supset d \left(\bar{\nu}_{L}  \sigma_{\mu \nu} F^{\mu \nu} N \right)$, dominates the phenomenology of the heavy singlet lepton $N$, but have not commented on the implications of such an operator.  As discussed in~\cite{Magill:2018jla}, if $N$ has a large Majorana mass the NDP interaction can generate a large active neutrino mass. However, this concern is mitigated if the Dirac mass is large compared to the Majorana mass. Moreover, it has been known that the Zee model~\cite{Zee:1980ai} of neutrino masses can naturally accommodate large magnetic moments~\cite{Babu:1992vq}. Other models that have been shown to allow for large magnetic moments consistently are the Barr, Freire, Zee (BFZ) spin-suppression mechanism~\cite{Barr:1990um} and horizontal flavor symmetries~\cite{Barbieri:1988fh} (see also~\cite{Lindner:2017uvt}).  }

\section{Conclusion}
\label{sec:conc}
We have shown that the recent XENON1T data can be used to constrain the neutrino dipole portal at new levels of sensitivity. The solar neutrinos which are nearly detectable at direct detection experiments would already have been observed if the NDP interaction was sufficiently large. Future improvements in searches such as these will come from large scale, low-threshold experiments which will be capable of seeing large numbers of solar neutrino events. We note that although we have focused on nuclear recoil events here, one could extend the analysis to include electron recoil events. Given the reduction in the CM energy from scattering on electrons, this method will only allow sensitivity to sterile masses $m_{4} \lesssim 0.5$ MeV. 
\\

{Lastly, we note that our computation of the bounds on the NDP coupling from direct detection experiments has been intentionally conservative. This is due to the imposition of the penalty factor Eq.\ref{eq:penalty} which vetoes events in which the singlet fermion decay $\nu_{4} \rightarrow \nu+\gamma$ occurs within the detector volume, and may therefore complicate discrimination from $e/\gamma$ backgrounds since it yields a nonstandard $S1$ and $S2$ signal in XENON1T. A similar effect arising from DM-induced nuclear excitations which re-decay to photon final states on short timescales was studied in~\cite{McCabe:2015eia}. A future analysis may be able to conduct a dedicated search for the non-standard $S1$ and $S2$ signals associated with the NDP, resulting in a strengthening of the bounds found here. We leave such a study to future work. }

\section*{Acknowledgments}
We are very grateful to Laura Baudis, Dongming Mei, Ryan Plestid, and Yu-Dai Tsai for helpful discussions. This work in supported by the U.S. Department of Energy under the award number DE-SC0019163.

\bibliographystyle{JHEP}

\bibliography{nu}

\providecommand{\href}[2]{#2}\begingroup\raggedright\begin{thebibliography}{10}

\bibitem{Gninenko:2009ks}
S.~N. Gninenko, \emph{{The MiniBooNE anomaly and heavy neutrino decay}},
  \href{https://doi.org/10.1103/PhysRevLett.103.241802}{\emph{Phys. Rev. Lett.}
  {\bfseries 103} (2009) 241802}
  [\href{https://arxiv.org/abs/0902.3802}{{\ttfamily 0902.3802}}].

\bibitem{Gninenko:2010pr}
S.~N. Gninenko, \emph{{A resolution of puzzles from the LSND, KARMEN, and
  MiniBooNE experiments}},
  \href{https://doi.org/10.1103/PhysRevD.83.015015}{\emph{Phys. Rev.}
  {\bfseries D83} (2011) 015015}
  [\href{https://arxiv.org/abs/1009.5536}{{\ttfamily 1009.5536}}].

\bibitem{McKeen:2010rx}
D.~McKeen and M.~Pospelov, \emph{{Muon Capture Constraints on Sterile Neutrino
  Properties}}, \href{https://doi.org/10.1103/PhysRevD.82.113018}{\emph{Phys.
  Rev.} {\bfseries D82} (2010) 113018}
  [\href{https://arxiv.org/abs/1011.3046}{{\ttfamily 1011.3046}}].

\bibitem{Masip:2011qb}
M.~Masip and P.~Masjuan, \emph{{Heavy-neutrino decays at neutrino telescopes}},
  \href{https://doi.org/10.1103/PhysRevD.83.091301}{\emph{Phys. Rev.}
  {\bfseries D83} (2011) 091301}
  [\href{https://arxiv.org/abs/1103.0689}{{\ttfamily 1103.0689}}].

\bibitem{Gninenko:2012rw}
S.~N. Gninenko, \emph{{New limits on radiative sterile neutrino decays from a
  search for single photons in neutrino interactions}},
  \href{https://doi.org/10.1016/j.physletb.2012.02.071}{\emph{Phys. Lett.}
  {\bfseries B710} (2012) 86}
  [\href{https://arxiv.org/abs/1201.5194}{{\ttfamily 1201.5194}}].

\bibitem{Masip:2012ke}
M.~Masip, P.~Masjuan and D.~Meloni, \emph{{Heavy neutrino decays at
  MiniBooNE}}, \href{https://doi.org/10.1007/JHEP01(2013)106}{\emph{JHEP}
  {\bfseries 01} (2013) 106} [\href{https://arxiv.org/abs/1210.1519}{{\ttfamily
  1210.1519}}].

\bibitem{Bertuzzo:2018itn}
E.~Bertuzzo, S.~Jana, P.~A.~N. Machado and R.~Zukanovich~Funchal, \emph{{A Dark
  Neutrino Portal to Explain MiniBooNE}},
  \href{https://arxiv.org/abs/1807.09877}{{\ttfamily 1807.09877}}.

\bibitem{Coloma:2017ppo}
P.~Coloma, P.~A.~N. Machado, I.~Martinez-Soler and I.~M. Shoemaker,
  \emph{{Double-Cascade Events from New Physics in Icecube}},
  \href{https://doi.org/10.1103/PhysRevLett.119.201804}{\emph{Phys. Rev. Lett.}
  {\bfseries 119} (2017) 201804}
  [\href{https://arxiv.org/abs/1707.08573}{{\ttfamily 1707.08573}}].

\bibitem{Magill:2018jla}
G.~Magill, R.~Plestid, M.~Pospelov and Y.-D. Tsai, \emph{{Dipole portal to
  heavy neutral leptons}},  \href{https://arxiv.org/abs/1803.03262}{{\ttfamily
  1803.03262}}.

\bibitem{Aprile:2018dbl}
{\scshape XENON} collaboration, E.~Aprile et~al., \emph{{Dark Matter Search
  Results from a One Tonne$\times$Year Exposure of XENON1T}},
  \href{https://arxiv.org/abs/1805.12562}{{\ttfamily 1805.12562}}.

\bibitem{Pospelov:2011ha}
M.~Pospelov, \emph{{Neutrino Physics with Dark Matter Experiments and the
  Signature of New Baryonic Neutral Currents}},
  \href{https://doi.org/10.1103/PhysRevD.84.085008}{\emph{Phys.Rev.} {\bfseries
  D84} (2011) 085008} [\href{https://arxiv.org/abs/1103.3261}{{\ttfamily
  1103.3261}}].

\bibitem{Harnik:2012ni}
R.~Harnik, J.~Kopp and P.~A.~N. Machado, \emph{{Exploring nu Signals in Dark
  Matter Detectors}},
  \href{https://doi.org/10.1088/1475-7516/2012/07/026}{\emph{JCAP} {\bfseries
  1207} (2012) 026} [\href{https://arxiv.org/abs/1202.6073}{{\ttfamily
  1202.6073}}].

\bibitem{Pospelov:2012gm}
M.~Pospelov and J.~Pradler, \emph{{Elastic scattering signals of solar
  neutrinos with enhanced baryonic currents}},
  \href{https://doi.org/10.1103/PhysRevD.85.113016,
  10.1103/PhysRevD.88.039904}{\emph{Phys. Rev.} {\bfseries D85} (2012) 113016}
  [\href{https://arxiv.org/abs/1203.0545}{{\ttfamily 1203.0545}}].

\bibitem{Pospelov:2013rha}
M.~Pospelov and J.~Pradler, \emph{{Dark Matter or Neutrino recoil?
  Interpretation of Recent Experimental Results}},
  \href{https://doi.org/10.1103/PhysRevD.89.055012}{\emph{Phys. Rev.}
  {\bfseries D89} (2014) 055012}
  [\href{https://arxiv.org/abs/1311.5764}{{\ttfamily 1311.5764}}].

\bibitem{Coloma:2014hka}
P.~Coloma, P.~Huber and J.~M. Link, \emph{{Combining dark matter detectors and
  electron-capture sources to hunt for new physics in the neutrino sector}},
  \href{https://doi.org/10.1007/JHEP11(2014)042}{\emph{JHEP} {\bfseries 11}
  (2014) 042} [\href{https://arxiv.org/abs/1406.4914}{{\ttfamily 1406.4914}}].

\bibitem{Cerdeno:2016sfi}
D.~G. Cerdeo, M.~Fairbairn, T.~Jubb, P.~A.~N. Machado, A.~C. Vincent and
  C.~Bhm, \emph{{Physics from solar neutrinos in dark matter direct detection
  experiments}}, \href{https://doi.org/10.1007/JHEP09(2016)048,
  10.1007/JHEP05(2016)118}{\emph{JHEP} {\bfseries 05} (2016) 118}
  [\href{https://arxiv.org/abs/1604.01025}{{\ttfamily 1604.01025}}].

\bibitem{Dent:2016wcr}
J.~B. Dent, B.~Dutta, S.~Liao, J.~L. Newstead, L.~E. Strigari and J.~W. Walker,
  \emph{{Probing light mediators at ultralow threshold energies with coherent
  elastic neutrino-nucleus scattering}},
  \href{https://doi.org/10.1103/PhysRevD.96.095007}{\emph{Phys. Rev.}
  {\bfseries D96} (2017) 095007}
  [\href{https://arxiv.org/abs/1612.06350}{{\ttfamily 1612.06350}}].

\bibitem{Bertuzzo:2017tuf}
E.~Bertuzzo, F.~F. Deppisch, S.~Kulkarni, Y.~F. Perez~Gonzalez and
  R.~Zukanovich~Funchal, \emph{{Dark Matter and Exotic Neutrino Interactions in
  Direct Detection Searches}},
  \href{https://doi.org/10.1007/JHEP04(2017)073}{\emph{JHEP} {\bfseries 04}
  (2017) 073} [\href{https://arxiv.org/abs/1701.07443}{{\ttfamily
  1701.07443}}].

\bibitem{Dutta:2017nht}
B.~Dutta, S.~Liao, L.~E. Strigari and J.~W. Walker, \emph{{Non-standard
  interactions of solar neutrinos in dark matter experiments}},
  \href{https://doi.org/10.1016/j.physletb.2017.08.031}{\emph{Phys. Lett.}
  {\bfseries B773} (2017) 242}
  [\href{https://arxiv.org/abs/1705.00661}{{\ttfamily 1705.00661}}].

\bibitem{AristizabalSierra:2017joc}
D.~Aristizabal~Sierra, N.~Rojas and M.~H.~G. Tytgat, \emph{{Neutrino
  non-standard interactions and dark matter searches with multi-ton scale
  detectors}}, \href{https://doi.org/10.1007/JHEP03(2018)197}{\emph{JHEP}
  {\bfseries 03} (2018) 197}
  [\href{https://arxiv.org/abs/1712.09667}{{\ttfamily 1712.09667}}].

\bibitem{Gonzalez-Garcia:2018dep}
M.~C. Gonzalez-Garcia, M.~Maltoni, Y.~F. Perez-Gonzalez and
  R.~Zukanovich~Funchal, \emph{{Neutrino Discovery Limit of Dark Matter Direct
  Detection Experiments in the Presence of Non-Standard Interactions}},
  \href{https://doi.org/10.1007/JHEP07(2018)019}{\emph{JHEP} {\bfseries 07}
  (2018) 019} [\href{https://arxiv.org/abs/1803.03650}{{\ttfamily
  1803.03650}}].

\bibitem{Boehm:2018sux}
C.~B{\oe}hm, D.~G. Cerde{\~n}o, P.~A.~N. Machado, A.~O.-D. Campo and E.~Reid,
  \emph{{How high is the neutrino floor?}},
  \href{https://arxiv.org/abs/1809.06385}{{\ttfamily 1809.06385}}.

\bibitem{Schwienhorst:2001sj}
{\scshape DONUT} collaboration, R.~Schwienhorst et~al., \emph{{A New upper
  limit for the tau - neutrino magnetic moment}},
  \href{https://doi.org/10.1016/S0370-2693(01)00746-8}{\emph{Phys. Lett.}
  {\bfseries B513} (2001) 23}
  [\href{https://arxiv.org/abs/hep-ex/0102026}{{\ttfamily hep-ex/0102026}}].

\bibitem{Cabrera:1984rr}
B.~Cabrera, L.~M. Krauss and F.~Wilczek, \emph{{Bolometric Detection of
  Neutrinos}}, \href{https://doi.org/10.1103/PhysRevLett.55.25}{\emph{Phys.
  Rev. Lett.} {\bfseries 55} (1985) 25}.

\bibitem{Monroe:2007xp}
J.~Monroe and P.~Fisher, \emph{{Neutrino Backgrounds to Dark Matter Searches}},
  \href{https://doi.org/10.1103/PhysRevD.76.033007}{\emph{Phys. Rev.}
  {\bfseries D76} (2007) 033007}
  [\href{https://arxiv.org/abs/0706.3019}{{\ttfamily 0706.3019}}].

\bibitem{Strigari:2009bq}
L.~E. Strigari, \emph{{Neutrino Coherent Scattering Rates at Direct Dark Matter
  Detectors}}, \href{https://doi.org/10.1088/1367-2630/11/10/105011}{\emph{New
  J. Phys.} {\bfseries 11} (2009) 105011}
  [\href{https://arxiv.org/abs/0903.3630}{{\ttfamily 0903.3630}}].

\bibitem{Gutlein:2010tq}
A.~Gutlein et~al., \emph{{Solar and atmospheric neutrinos: Background sources
  for the direct dark matter search}},
  \href{https://doi.org/10.1016/j.astropartphys.2010.06.002}{\emph{Astropart.
  Phys.} {\bfseries 34} (2010) 90}
  [\href{https://arxiv.org/abs/1003.5530}{{\ttfamily 1003.5530}}].

\bibitem{Billard:2013qya}
J.~Billard, L.~Strigari and E.~Figueroa-Feliciano, \emph{{Implication of
  neutrino backgrounds on the reach of next generation dark matter direct
  detection experiments}},
  \href{https://doi.org/10.1103/PhysRevD.89.023524}{\emph{Phys. Rev.}
  {\bfseries D89} (2014) 023524}
  [\href{https://arxiv.org/abs/1307.5458}{{\ttfamily 1307.5458}}].

\bibitem{Ruppin:2014bra}
F.~Ruppin, J.~Billard, E.~Figueroa-Feliciano and L.~Strigari,
  \emph{{Complementarity of dark matter detectors in light of the neutrino
  background}}, \href{https://doi.org/10.1103/PhysRevD.90.083510}{\emph{Phys.
  Rev.} {\bfseries D90} (2014) 083510}
  [\href{https://arxiv.org/abs/1408.3581}{{\ttfamily 1408.3581}}].

\bibitem{Dent:2016iht}
J.~B. Dent, B.~Dutta, J.~L. Newstead and L.~E. Strigari, \emph{{Effective field
  theory treatment of the neutrino background in direct dark matter detection
  experiments}}, \href{https://doi.org/10.1103/PhysRevD.93.075018}{\emph{Phys.
  Rev.} {\bfseries D93} (2016) 075018}
  [\href{https://arxiv.org/abs/1602.05300}{{\ttfamily 1602.05300}}].

\bibitem{OHare:2016pjy}
C.~A. O'Hare, \emph{{Dark matter astrophysical uncertainties and the neutrino
  floor}}, \href{https://doi.org/10.1103/PhysRevD.94.063527}{\emph{Phys. Rev.}
  {\bfseries D94} (2016) 063527}
  [\href{https://arxiv.org/abs/1604.03858}{{\ttfamily 1604.03858}}].

\bibitem{Essig:2018tss}
R.~Essig, M.~Sholapurkar and T.-T. Yu, \emph{{Solar Neutrinos as a Signal and
  Background in Direct-Detection Experiments Searching for Sub-GeV Dark Matter
  With Electron Recoils}},
  \href{https://doi.org/10.1103/PhysRevD.97.095029}{\emph{Phys. Rev.}
  {\bfseries D97} (2018) 095029}
  [\href{https://arxiv.org/abs/1801.10159}{{\ttfamily 1801.10159}}].

\bibitem{Wyenberg:2018eyv}
J.~Wyenberg and I.~M. Shoemaker, \emph{{Mapping the neutrino floor for direct
  detection experiments based on dark matter-electron scattering}},
  \href{https://doi.org/10.1103/PhysRevD.97.115026}{\emph{Phys. Rev.}
  {\bfseries D97} (2018) 115026}
  [\href{https://arxiv.org/abs/1803.08146}{{\ttfamily 1803.08146}}].

\bibitem{Goodman:1984dc}
M.~W. Goodman and E.~Witten, \emph{{Detectability of Certain Dark Matter
  Candidates}}, \href{https://doi.org/10.1103/PhysRevD.31.3059}{\emph{Phys.
  Rev.} {\bfseries D31} (1985) 3059}.

\bibitem{Drukier:1983gj}
A.~Drukier and L.~Stodolsky, \emph{{Principles and Applications of a Neutral
  Current Detector for Neutrino Physics and Astronomy}},
  \href{https://doi.org/10.1103/PhysRevD.30.2295}{\emph{Phys. Rev.} {\bfseries
  D30} (1984) 2295}.

\bibitem{Abe:2010hy}
{\scshape Super-Kamiokande} collaboration, K.~Abe et~al., \emph{{Solar neutrino
  results in Super-Kamiokande-III}},
  \href{https://doi.org/10.1103/PhysRevD.83.052010}{\emph{Phys. Rev.}
  {\bfseries D83} (2011) 052010}
  [\href{https://arxiv.org/abs/1010.0118}{{\ttfamily 1010.0118}}].

\bibitem{Akerib:2016mzi}
{\scshape LUX} collaboration, D.~S. Akerib et~al., \emph{{Low-energy (0.7-74
  keV) nuclear recoil calibration of the LUX dark matter experiment using D-D
  neutron scattering kinematics}},
  \href{https://arxiv.org/abs/1608.05381}{{\ttfamily 1608.05381}}.

\bibitem{Aprile:2017iyp}
{\scshape XENON} collaboration, E.~Aprile et~al., \emph{{First Dark Matter
  Search Results from the XENON1T Experiment}},
  \href{https://doi.org/10.1103/PhysRevLett.119.181301}{\emph{Phys. Rev. Lett.}
  {\bfseries 119} (2017) 181301}
  [\href{https://arxiv.org/abs/1705.06655}{{\ttfamily 1705.06655}}].

\bibitem{Tanabashi:2018oca}
{\scshape Particle Data Group} collaboration, M.~Tanabashi et~al.,
  \emph{{Review of Particle Physics}},
  \href{https://doi.org/10.1103/PhysRevD.98.030001}{\emph{Phys. Rev.}
  {\bfseries D98} (2018) 030001}.

\bibitem{Aprile:2011hx}
{\scshape XENON100} collaboration, E.~Aprile et~al., \emph{{Likelihood Approach
  to the First Dark Matter Results from XENON100}},
  \href{https://doi.org/10.1103/PhysRevD.84.052003}{\emph{Phys. Rev.}
  {\bfseries D84} (2011) 052003}
  [\href{https://arxiv.org/abs/1103.0303}{{\ttfamily 1103.0303}}].

\bibitem{Altegoer:1997gv}
{\scshape NOMAD} collaboration, J.~Altegoer et~al., \emph{{The NOMAD experiment
  at the CERN SPS}},
  \href{https://doi.org/10.1016/S0168-9002(97)01079-6}{\emph{Nucl. Instrum.
  Meth.} {\bfseries A404} (1998) 96}.

\bibitem{Geiregat:1989sz}
{\scshape CHARM-II} collaboration, D.~Geiregat et~al., \emph{{A New
  Determination of the Electroweak Mixing Angle From $\nu_\mu$ Electron
  Scattering}}, \href{https://doi.org/10.1016/0370-2693(89)90457-7}{\emph{Phys.
  Lett.} {\bfseries B232} (1989) 539}.

\bibitem{AguilarArevalo:2007it}
{\scshape MiniBooNE} collaboration, A.~A. Aguilar-Arevalo et~al., \emph{{A
  Search for electron neutrino appearance at the $\Delta m^{2} \sim 1$eV$^{2}$
  scale}}, \href{https://doi.org/10.1103/PhysRevLett.98.231801}{\emph{Phys.
  Rev. Lett.} {\bfseries 98} (2007) 231801}
  [\href{https://arxiv.org/abs/0704.1500}{{\ttfamily 0704.1500}}].

\bibitem{Agnese:2018col}
{\scshape SuperCDMS} collaboration, R.~Agnese et~al., \emph{{First Dark Matter
  Constraints from a SuperCDMS Single-Charge Sensitive Detector}},
  \href{https://doi.org/10.1103/PhysRevLett.121.051301}{\emph{Phys. Rev. Lett.}
  {\bfseries 121} (2018) 051301}
  [\href{https://arxiv.org/abs/1804.10697}{{\ttfamily 1804.10697}}].

\bibitem{Agnese:2017jvy}
{\scshape SuperCDMS} collaboration, R.~Agnese et~al., \emph{{Low-mass dark
  matter search with CDMSlite}},
  \href{https://doi.org/10.1103/PhysRevD.97.022002}{\emph{Phys. Rev.}
  {\bfseries D97} (2018) 022002}
  [\href{https://arxiv.org/abs/1707.01632}{{\ttfamily 1707.01632}}].

\bibitem{Zee:1980ai}
A.~Zee, \emph{{A Theory of Lepton Number Violation, Neutrino Majorana Mass, and
  Oscillation}}, \href{https://doi.org/10.1016/0370-2693(80)90349-4,
  10.1016/0370-2693(80)90193-8}{\emph{Phys. Lett.} {\bfseries 93B} (1980) 389}.

\bibitem{Babu:1992vq}
K.~S. Babu, D.~Chang, W.-Y. Keung and I.~Phillips, \emph{{Comment on `Mechanism
  for large neutrino magnetic moments'}},
  \href{https://doi.org/10.1103/PhysRevD.46.2268}{\emph{Phys. Rev.} {\bfseries
  D46} (1992) 2268}.

\bibitem{Barr:1990um}
S.~M. Barr, E.~M. Freire and A.~Zee, \emph{{A Mechanism for large neutrino
  magnetic moments}},
  \href{https://doi.org/10.1103/PhysRevLett.65.2626}{\emph{Phys. Rev. Lett.}
  {\bfseries 65} (1990) 2626}.

\bibitem{Barbieri:1988fh}
R.~Barbieri and R.~N. Mohapatra, \emph{{A Neutrino With a Large Magnetic Moment
  and a Naturally Small Mass}},
  \href{https://doi.org/10.1016/0370-2693(89)91423-8}{\emph{Phys. Lett.}
  {\bfseries B218} (1989) 225}.

\bibitem{Lindner:2017uvt}
M.~Lindner, B.~Radovi and J.~Welter, \emph{{Revisiting Large Neutrino Magnetic
  Moments}}, \href{https://doi.org/10.1007/JHEP07(2017)139}{\emph{JHEP}
  {\bfseries 07} (2017) 139}
  [\href{https://arxiv.org/abs/1706.02555}{{\ttfamily 1706.02555}}].

\bibitem{McCabe:2015eia}
C.~McCabe, \emph{{Prospects for dark matter detection with inelastic
  transitions of xenon}},
  \href{https://doi.org/10.1088/1475-7516/2016/05/033}{\emph{JCAP} {\bfseries
  1605} (2016) 033} [\href{https://arxiv.org/abs/1512.00460}{{\ttfamily
  1512.00460}}].

\end{thebibliography}\endgroup

\end{document}